\documentclass[twocolumn,aps,floatfix,showpacs]{revtex4}
\newcommand{\be}{\begin{equation}}
\newcommand{\ee}{\end{equation}}
\newcommand{\bea}{\begin{eqnarray}}
\newcommand{\eea}{\end{eqnarray}}

\usepackage{graphicx}
\usepackage{bm}
\begin{document}

\title{Feshbach resonances in ultracold $^{39}$K}
\author{Chiara D'Errico, Matteo Zaccanti$^1$, Marco Fattori $^{3}$, Giacomo Roati$^{1,2}$, Massimo Inguscio$^{1,2}$, Giovanni Modugno$^{1,2}$}
\affiliation{LENS and Dipartimento di Fisica, Universit\`a di Firenze,Via Nello Carrara 1, 50019 Sesto Fiorentino,
Italy \\
$^{1}$ INFN, Via Sansone 1, 50019 Sesto Fiorentino, Italy \\
$^{2}$ CNR-INFM, Via Sansone 1, 50019 Sesto Fiorentino, Italy \\
$^{3}$ Museo Storico della Fisica e Centro Studi e Ricerche "Enrico Fermi", Piazzale del Viminale 2 - Compendio del Viminale Pal.F, 00184 Roma, Italy}
\author{Andrea Simoni}
\affiliation{Laboratoire de Physique des Atomes, Lasers, Mol\'ecules et Surfaces \\ UMR 6627 du CNRS and Universit\'e de Rennes, 35042 Rennes Cedex, France} 
\begin{abstract}
We discover several magnetic Feshbach resonances in collisions of
ultracold $^{39}$K atoms, by studying atom losses and molecule
formation. Accurate determination of the magnetic-field resonance
locations allows us to optimize a quantum collision model for
potassium isotopes. We employ the model to predict the
magnetic-field dependence of scattering lengths and of
near-threshold molecular levels. Our findings will be useful to plan
future experiments on ultracold potassium atoms and molecules.

\end{abstract}

\date{\today}
\maketitle
\section{Introduction}
Control of the isotropic interaction in ultracold atomic gases
\cite{feshbach,sodium} is revealing itself as a fundamental tool to
explore a variety of fundamental phenomena. Tuning the interaction
between two different hyperfine states in Fermi gases via magnetic
Feshbach resonances has permitted unprecedented investigations of
the BEC-BCS crossover \cite{becbcs}. Mean-field effects such as
collapse \cite{collapse} or formation of bright solitons
\cite{solitons} have been demonstrated in Bose-Einstein condensates
with tunable interactions. Homonuclear Feshbach resonances have also
been successfully used to convert atomic gases in molecular
Bose-Einstein condensates \cite{molbec}, to produce strongly
correlated quantum phases \cite{mott,repulsive} and to observe
Efimov trimer states \cite{efimov}. Analogous experiments with
heteronuclear systems are in progress \cite{hetero}.

Feshbach resonances have been discovered in most alkali species,
including Li \cite{lithium}, Na \cite{sodium}, K \cite{potassium} Rb
\cite{rubidium}, Cs\cite{cesium}, in chromium \cite{chromium} and in
few alkali mixtures \cite{mixtures}. In the case of potassium,
intensive study of two specific resonances in fermionic $^{40}$K
\cite{potassium} has been motivated by possible applications
to fermionic superfluidity. Although ultracold samples of the
bosonic isotopes $^{41}$K and $^{39}$K have also been produced
\cite{modugno,desarlo}, magnetic Feshbach resonances have not yet
been investigated in these systems. Moreover, only limited
theoretical predictions exist for these isotopes
\cite{old,tiesinga}.

We report here the first experimental study of Feshbach resonances
in an ultracold $^{39}$K gas. We discover several resonances in
three different hyperfine states and measure their magnetic-field
location by observing on-resonance enhancement of inelastic
three-body losses and molecule formation. Each hyperfine state of
interest presents at least one broad Feshbach resonance which can be
used to tune with high accuracy the interaction in a $^{39}$K
Bose-Einstein condensate \cite{nostro}.

The observed resonance locations are used to construct an accurate
theoretical quantum model which explains both present and
pre-existing observations \cite{potassium}. The model allows us to
compute relevant quantities such as background scattering lengths
and resonance widths. In addition, we fully characterize
hyperfine-coupled molecular levels near the dissociation limit.
Knowledge of molecular parameters is essential for understanding
experiments performed in the strongly interacting regime. It is also
important for implementing schemes of molecules formation and for
the interpretation of their properties.

\section{Experiment} The apparatus and techniques we use
to prepare ultracold samples of $^{39}$K atoms have already been
presented in detail elsewhere \cite{nostro} and are only briefly
summarized here. We begin by preparing a mixture of $^{39}$K and
$^{87}$Rb atoms in a magneto-optical trap at temperatures of the
order of few 100 $\mu$K. We simultaneously load the two species in a
magnetic potential in their stretched Zeeman states $|F=2,
m_F=2\rangle$ and perform 25 s of selective evaporation of rubidium
on the hyperfine transition at 6.834 GHz. Potassium is
sympathetically cooled through interspecies collisions
\cite{desarlo}. When the mixture temperature is below 1$\mu$K it is
transferred to an optical potential. This is created with
two focused laser beams at a wavelength $\lambda$=1030 nm, with beam
waists of about 100 $\mu$m and crossing in the horizontal plane.

\begin{figure}[htbp]
\includegraphics[width=\columnwidth,clip]{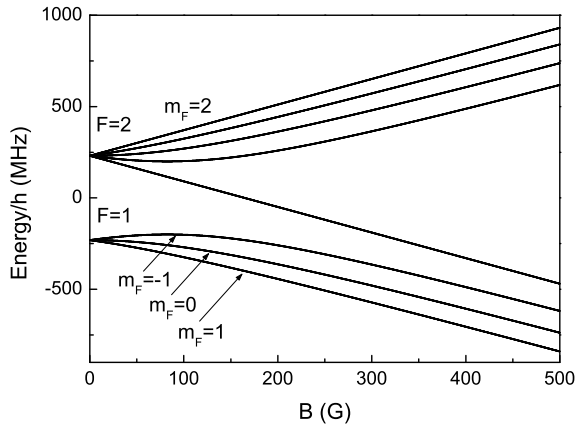}
\caption{Energy levels of $^{39}$K atoms in a magnetic field.
Levels are labeled using zero-field quantum numbers.}
\label{fig0}
\end{figure}

In this work we have studied Feshbach resonances in all the states
immune from spin-exchange collisions, the three Zeeman sublevels of
the $F$=1 manifold \cite{inn}. The level structure of $^{39}$K in a
magnetic field is shwon in Fig. \ref{fig0}. The atoms are initially
prepared in the $|1, 1\rangle$ state by adiabatic rapid passage over
the hyperfine transition around 462 MHz \cite{violino} in a 10~G
homogeneous magnetic field. To further cool the sample, we transfer
also Rb to its ground state, and we lower the optical trap depth by
exponentially decreasing the laser power in 2.4 s. During the forced
evaporation of both species we increase the K-Rb elastic cross
section by applying a homogeneous magnetic field of 316 G, close to
an interspecies Feshbach resonance \cite{nostro}. With this
technique we are able to cool the K sample to a final temperature in
the range 150-500 nK. At these temperatures the sample is not yet
quantum degenerate. Once the K sample has been prepared to the
desired temperature, Rb is selectively removed from the trap using a
resonant light pulse. To transfer the atoms in the two excited
states of the $F$=1 manifold, we use a radio-frequency sweep. For
the transfer from $|1, 1\rangle$ to $|1, -1\rangle$ we apply a 10~G
field and use a radio-frequency sweep about 7.6 MHz. For the
transfer from $|1, 1\rangle$ to $|1, 0\rangle$ we use instead a 38.5
G field and a radio-frequency sweep around 28.5 MHz. After the atoms
have been prepared in a given state (this typically requires from 10
to 30 ms, depending on the final state) we change the homogeneous
field in few ms and actively stabilize it to any desired value in
the 0-1000~G range with an accuracy of about 100~mG. We calibrate
the field by means of microwave and RF spectroscopy on two different
hyperfine transitions of Rb.

\begin{figure}[htbp]
\includegraphics[width=\columnwidth,clip]{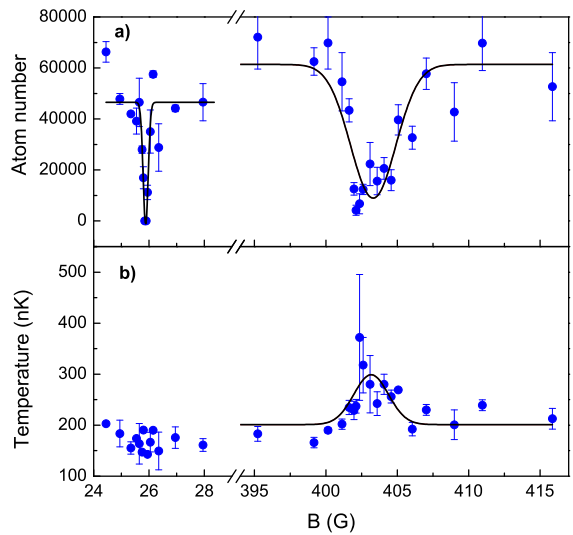}
\caption{Experimental determination of Feshbach resonances in the
$|1,1\rangle$ state of $^{39}$K: a) remaining atom number; b) sample
temperature. The hold time for the low field (high field) resonance
was 480 ms (36 ms). The curves are phenomenological fits with
gaussian distributions.} \label{fig1}
\end{figure}

Feshbach resonances are detected as an enhancement of losses. In
proximity of a Feshbach resonance, the scattering length can be
parametrized as \cite{feshbach}
\begin{equation}
a(B)=a_{\rm bg}\left( 1-\frac{\Delta}{B-B_0} \right) \,,
\label{isofesh}
\end{equation}
where $a_{\rm bg}$ is the background scattering length, $\Delta$ is
the resonance width, and $B_0$ the resonance center. The resonance
width $\Delta$ represents the separation between $B_0$ and the
location where the scattering length crosses zero. As $a(B)$
diverges, two- and three-body inelastic rates are enhanced,
resulting in atom loss from the trap and heating \cite{threebody}.
This potassium isotope presents both very narrow ($\Delta <$0.5 G)
and very broad resonances ($\Delta \sim$50 G). Examples of two such
resonances in the $|1, 1\rangle$ state are shown in Fig. \ref{fig1}.
In the present experiment the magnetic field was brought to a
variable value in the range 0-500 G, and held there for a given
time. Both field and trapping laser were then switched off and atom
number and temperature were measured through standard absorption
imaging. The narrow resonance around 26 G gives rise to a rather
sharp, symmetric loss features centered at $B_0$. On the converse,
the broad resonance around 400 G corresponds with broader, highly
asymmetric loss and heating features. A possible source of asymmetry
will be discussed later. The different strength of the two
resonances is indicated by the different hold time required to have
about 90\% peak losses; this amounts to 480 ms for the narrow
resonance \cite{note heating} and 36 ms for the broad one.

\begin{table*}[t]
\begin{center}
\caption{Experimental magnetic-field positions $B_{\rm exp}$ and
theoretically calculated positions $B_{\rm th}$, widths $\Delta$,
magnetic moments $s$, background scattering length $a_{\rm bg}$, and
approximate quantum numbers (see text) of $^{39}$K $\ell=0$ Feshbach
resonances. } \label{table1} \vskip 12pt
\begin{tabular}{l | c c c cc c c}
\hline \hline
 $m_{F},m_{F}$& $B_{\rm exp}$ (G)& $B_{\rm th}$(G)&$-\Delta_{\rm th}$ (G)
 & -$s$~($\mu_{\rm B}$) & $a_{\rm bg}(a_0)$ & $(SIf)$ or $\{ SIM_S\}$\\
 \hline \hline
$1,1$ &25.85(10)& 25.9  & 0.47 & 1.5  & -33 & $(133)$\\
                        &403.4(7) &402.4  & 52 & 1.5  & -29 & \\ 
                        &         &745.1  & 0.4  & 3.9  & -35 & $\{113\}$  \\
                        &752.3(1)         &752.4  & 0.4  & 3.9  & -35 & $\{ 111\}$\\
 \hline
$0,0$                   &59.3(6)   &58.8   & 9.6  & 0.83 & -18 & $(133)$\\
                        &66.0(9)   &65.6   & 7.9  & 0.78 & -18 & $(111)$   \\
                        &          &471    & 72   & 1.22 & -28 & \\ 
                        &          &490    &  5   & 1.70 & -28 & \\ 
                        &          &825    & 0.032& 3.92 & -36 & $\{ 1 1 3\}$ \\
                        &          &832    & 0.52 & 3.90 & -36 & $\{ 1 1 1\}$ \\
\hline
$-1,-1$                 &32.6(1.5)  &33.6        & -55  & -1.9 & -19 & $(112)$     \\
                        &162.8(9)   &162.3       & 37   &  1.2 & -19 & $(133)$     \\
                        &562.2(1.5) &560.7       & 56   &  1.4 & -29 & \\ 
 \hline \hline
\end{tabular}
\end{center}
\end{table*}

The same procedure was repeated for the two other hyperfine states.
In total we have studied eight Feshbach resonances, whose centers
are listed in Tab. \ref{table1}. For most broad resonances, we
have found an asymmetry in loss and heating profiles similar to the
one shown in Fig. \ref{fig1}. In the absence of a precise model of
our system, we have fitted the experimental profiles with a single
gaussian to determine the resonance centers $B_{\rm exp}$. The error
we give on $B_{\rm exp}$ is the quadratic sum of our magnetic-field
accuracy and of the error deriving from the fit, which is usually
dominating for broad loss profiles.

\section{Theoretical analysis}

Early information about K collision properties was obtained from the
analysis of photoassociation spectra of the bosonic isotope
$^{39}$K, see \cite{bohnburke,williams1u}. The collision model has
then been refined by theoretically analyzing observed shape
\cite{demarco} and Feshbach resonances \cite{potassium} in fermionic
$^{40}$K. Subsequently, the Nist/Connecticut groups have inferred
potential parameters from two-photon spectroscopy of $^{39}$K
near-dissociation molecular levels \cite{wang}. Finally, cold
collision measurements have been performed on $^{39}$K
\cite{desarlo}. These different determinations are summarized in
Tab.~\ref{table2} (scattering quantities are defined in the
following).

\begin{table}[t]
\begin{center}
\caption{Comparison of collisional parameters for $^{39}$K
determined from cold collisions (CC) and
photoassociation (PA) spectroscopy of ultracold atoms. Some analyses
did not determine the value of $C_6$, which was taken from theory
(value and reference are then reported in the third column of the table).
The $\delta C_6$ is the shift in $C_6$ from the value
$C_6=3897$~a.u. of Ref.~\cite{Derevianko}. } \label{table2} \vskip
12pt
\begin{tabular}{l |  c c c }
\hline \hline
{\rm Ref.}  & $a_S (a_0)$ & $a_T (a_0)$ & $C_6 (a.u.)$ \\
\hline
This work & $138.90 \pm 0.15$  &  $-33.3 \pm 0.3$  & $3921 \pm 8$   \\
CC \cite{desarlo} &   & $-51 \pm 7$  &    \\
CC \cite{potassium} & $139.4 \pm 0.7$ & $-37 \pm 6$ &  $ 3927 \pm 50 $     \\
PA \cite{wang} &             & $-33 \pm 5$  & $3897 \pm 15 $ \cite{Derevianko}   \\
CC \cite{demarco} &  & $>-80,<-28  $  &  $3813$ \cite{Marinescu}  \\
PA \cite{williams1u} & $>90,<230$ & $>-60,<15  $  & $3813$  \cite{Marinescu}         \\
PA \cite{bohnburke,old} & $140_{-9}^{+6}$ & $-21-0.045~\delta C_6 \pm 20$  &    \\

 \hline \hline
\end{tabular}
\end{center}
\end{table}

Our present collision model comprises adiabatic Born-Oppenheimer
singlet $X^1\Sigma^+$ and triplet $a^3\Sigma^+$ interaction
potentials determined from spectroscopic data
\cite{singlet,triplet}. The adiabatic potentials asymptotically
correlate with the well known dispersion plus exchange analytical
form \be V(^{1,3}\Sigma^+ )\to
-\frac{C_6}{R^6}-\frac{C_8}{R^8}-\frac{C_{10}}{R^{10}}\mp A R^\alpha
e^{-\beta R} , \ee where $\alpha=7/\beta-1 $, $\beta=\sqrt{8I} $,
$I$ is the atomic first ionization energy in hartrees \cite{smirnov}
and $A$ is a positive constant. Hyperfine interactions are mostly
important at large internuclear separation and are safely
approximated by their atomic limit. A short-range correction is
finally added to the adiabatic potentials to model the data
\cite{old}.

The experimental resonance locations are used in a {\it weighted}
least square procedure to determine the correction size. The
resulting optimized potentials are parametrized in terms of $s$-wave
singlet $a_S$ and triplet $a_T$ scattering lengths and of the
long-range parameters $C_n,n=6,8,10$. Resonance positions are mainly
sensitive to the leading van der Waals coefficient $C_6$, which
along with the $a_{S,T}$ is a a fitting parameter in our procedure.
In order to obtain maximum constraint we also include in the
empirical data the positions of two already known $^{40}$K
resonances~\cite{potassium}, and a $p$-wave resonance we have
recently discovered at $\sim 436$ G in collisions of $^{40}$K
$|9/2,7/2\rangle$ atoms. We use the same potential for the two
isotopes assuming thereby the validity of the Born-Oppenheimer
approximation. Result of the fit is: $a_S=(138.90 \pm 0.15) a_0$,
$a_T=(-33.3 \pm 0.3) a_0$, and $C_6=(3921 \pm 8)$~a.u. The final
reduced value is $\chi^2=0.52$ only. Our singlet-triplet
scattering lengths agree well with previous determinations (see
Tab.~\ref{table1}) and represent an improvement of more than one
order of magnitude in $a_T$. The $C_6$ agrees to one standard
deviation with the accurate value of Derevianko {\it et
al.}~\cite{Derevianko}, $C_6=3897 \pm 15$~a.u. The singlet-triplet
scattering lengths of $^{40}$K computed with the present model are
$104.56 \pm 0.10$ and $169.7 \pm 0.4$, in very good agreement with
Ref.~\cite{potassium}.

A magnetic Feshbach resonance arises at a value $B_0$ of the
magnetic field when the energy of the separated atom pair becomes degenerate
with the energy of a molecular bound level. Scattering near a
magnetic resonance is fully characterized \cite{goral} by assigning
$B_0$, $\Delta$, the background scattering length $a_{\rm bg}$, the
$C_6$ coefficient, the magnetic moment $s$ of the molecule
associated to the resonance with respect to free atoms \be
s=\frac{\partial ( E_{\rm at} - E_{\rm mol} ) }{\partial B}, \ee
where $E_{\rm at}$ and $E_{\rm mol}$ represent the energy of the
separated atoms and of the molecule, respectively, and the
derivative is taken away from resonance. Parameters values for
observed and theoretically predicted resonances are found in
Tab.~\ref{table1}.

In cases where resonances are overlapping (i.e. when the magnetic
width is comparable to their magnetic field separation) we will
parametrize the effective scattering length with one background
parameter $a_{\rm bg}$, two widths $\Delta_{i}$ and two positions
$B_{0,i}$ ($i=1,2$) as \be a(B)=a_{\rm bg}\left(
1-\frac{\Delta_1}{B-B_{0,1}}-\frac{\Delta_2}{B-B_{0,2}}  \right).
\ee This expression clearly reduces to Eq.~\ref{isofesh}
when the resonances are isolated, $|B_{0,2}-B_{0,1}| \gg \Delta_1,\Delta_2$.

\begin{figure}[htbp]
\includegraphics[width=\columnwidth,clip]{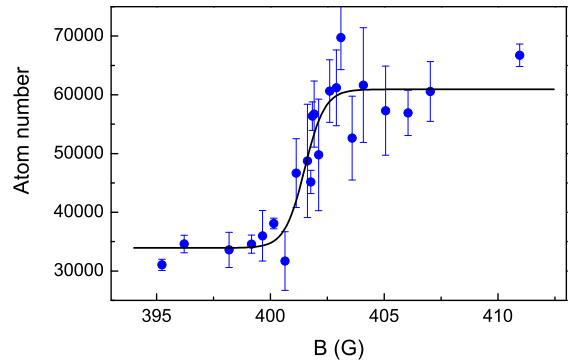}
\caption{Molecule formation at the broadest Feshbach resonances in
the $|1,1\rangle$ state of $^{39}$K. The magnetic field is linearly
swept from 410 G to a final field $B$ in 2 ms. The resonance center
$B_0$=401.5(5) G is determined by fitting the atom number with a
Boltzmann growth function.} \label{fig2}
\end{figure}

Comparison of experimental and theoretical resonance locations in
Tab. \ref{table1} indicates that all measured resonances with large
$\Delta$ feature an asymmetric profile. In all these cases, the
center of the gaussian fit to the loss profiles is indeed shifted
towards the region of negative scattering lengths, as in the case
reported in Fig. \ref{fig1}. A possible explanation for such
asymmetry are mean field effects for large positive and negative
scattering lengths close to the resonance center. For $B>B_0$ the
density should indeed increase with respect to the non-interacting
value, while for $B<B_0$ it should decrease. This would accordingly
vary the loss rates through their density dependencies and promote
losses on region with $B>B_0$. In absence of a detailed model of our
finite temperature system, we made an independent experiment to
determine the center of the broad ground-state resonance in Fig.
\ref{fig1}, by studying molecule association. We used the standard
technique of adiabatic magnetic-field sweeps over the resonance from
the atomic to the molecular side \cite{molbec}. The system was
initially prepared at a magnetic field well above the resonance
center, $B_{i}$=410 G, at a temperature of 220 nk. The field was
then swept to a final lower field $B$ in 2 ms, left to stabilize for
0.1 ms, then suddenly switched off. As shown in Fig.\ref{fig2}, as
$B$ crosses the resonance the atom number drops to about 50\% of the
initial value, in the absence of any heating of the system. This
indicates that a fraction of the atoms are converted into
weakly-bound molecules. The molecules are very rapidly lost from the
trap via inelastic collisions. A fit using a Boltzmann growth
function gives a resonance center of $B_0$=401.5(5) G. This is
almost 2 G lower than the center of the broad loss profile, and is
consistent with both $B_{\rm th}$=402.4(2) G and the value at which
the maximum atom loss and heating is seemingly taking place in the
data shown in Fig. \ref{fig1}, $B$=402.2(2) G. This agreement
confirms that the global fit we make is able to accurately fix the
position of all resonances, although the broad resonances centers
are individually determined with poorer accuracy by loss
measurements.

\begin{figure}
\includegraphics[width=\columnwidth,clip]{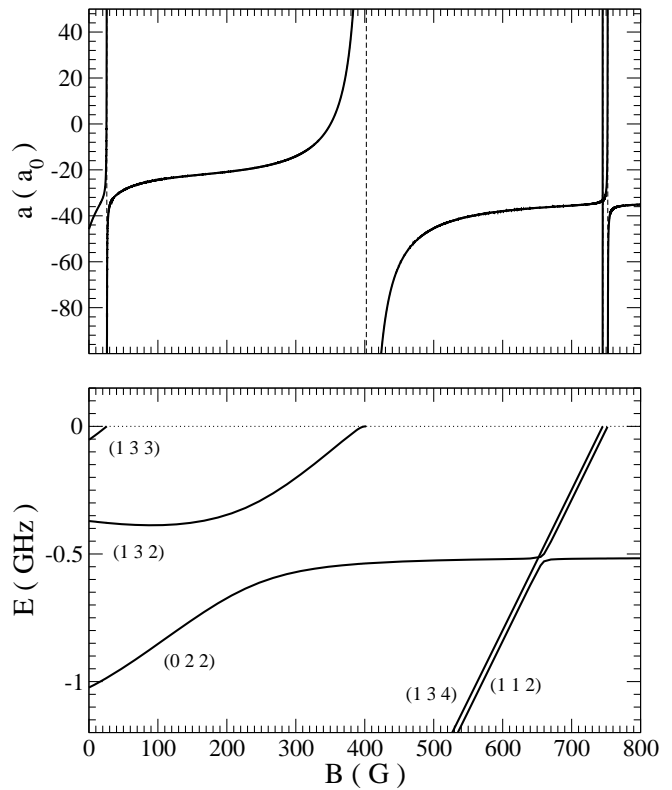}
\caption{Upper panel: magnetic field dependence of the effective
scattering length for $|1,1\rangle + |1,1\rangle$ $^{39}$K
collisions. Dashed lines indicate the resonance positions. Lower
panel: near-threshold molecular levels for $m_f=2$. Zero energy is taken at the
separated atoms limit. The quantum numbers shown in brackets $(SIf)$
are good in general only for weak magnetic fields, see text. }
\label{fig3}
\end{figure}

In order to complete the resonance characterization we now discuss
approximate quantum labels of the Feshbach molecule. Neglecting weak
dipolar interactions and for vanishing magnetic field the internal
angular momentum $\vec{f}=\vec{S}+\vec{I}$ is conserved. Here $\vec
S$ and $\vec I$ are the electronic and nuclear spin, respectively.
Moreover, because of the small hyperfine splitting of $^{39}$K with
respect to the splitting between neighbouring singlet-triplet
levels, $S$ and $I$ are approximately good quantum numbers at least
for low $B$. Because of the spherical symmetry of the problem, the
orbital angular momentum $\vec{\ell}$ of the atoms is also a
conserved quantity. All of our observed resonances have $\ell=0$.
Zero-energy quantum numbers are shown in Figs.
(\ref{fig3},\ref{fig4},\ref{fig5}) for the closest to dissociation
levels.

As the field increases these quantum numbers are not any longer
good. In fact, for intense magnetic fields the Zeeman energy becomes
larger than both the hyperfine and the singlet/triplet vibrational
splitting. In this regime $\vec S$ and $\vec I$ uncouple and precede
independently about the magnetic field. The molecular quantum state
can then be identified by $S,I$ and by the spin projections $M_S$
and $M_I$ on the quantization axis.

In the intermediate regime neither coupling scheme is accurate as
singlet and triplet levels are sufficiently close to be strongly
mixed by off-diagonal hyperfine interactions.
However, axial simmetry of the problem enforces conservation of the
magnetic quantum number $m_f$ (i.e. the axial projection of $\vec f$).
Examples of resonances
arising from such mixed levels are the $402$~G
(Fig.~\ref{fig3}), the  $471$~G and $490$~G
(Fig.~\ref{fig4}), and the $561$~G (Fig.~\ref{fig5}) features.
One can note from the figures broad avoided
crossings caused by spin-exchange interaction between levels of
different $S$ and same $f$. An approximate assignment constructed
for low and high field by averaging the appropriate spin operators
on the molecular wave functions is presented in Tab.~\ref{table1}.
Resonances arising from mixed levels are left unassigned. Their
zero-field correlation can be easily inferred from the figures.

One should also note that the quantum numbers discussed above are in
principle only valid away from resonance. Actually, there is always
a range of magnetic fields near resonance where the amplitude of the
molecular state is almost entirely transferred to the open
background channel~\cite{goral}, which is not represented by the
same quantum numbers as the molecule.

This magnetic field region can be
estimated as~\cite{petrov,goral} \be \frac{B-B_0}{\Delta} \ll
\frac{m a_{\rm bg}^2 s \Delta}{\hbar^2}, \label{broad} \ee with $m$
the atomic mass.
Resonances for
which the right hand side of Eq.~(\ref{broad}) is $\gg 1$ are termed
open channel dominated. The present resonances range from
closed channel dominated (${\rm rhs}\ll 1 $) to an intermediate situation (${\rm rhs} \simeq 1 $).
When condition (\ref{broad}) is fulfilled, the energy of the molecule
takes the form \bea E_{\rm mol}(B)- E_{\rm
at}(B)&=&-\frac{\hbar^2}{m \left[ a(B)-l_{\rm vdW}  \right]^2 } \nonumber \\
l_{\rm vdW}&=&\frac{1}{2}\left( \frac{ m C_6 }{\hbar^2} \right)^{1/4} \label{universal} \eea
see e.g.~\cite{goral,petrov}, and scattering can be described in terms
of a single effective channel. The $l_{\rm vdW}$ is the typical length
associated to a $R^{-6}$ interaction.
Validity of estimate (\ref{broad}) is confirmed by inspection of
Figs.(\ref{fig3},\ref{fig4},\ref{fig5}) in which the asymptotic
behaviour (\ref{universal}) is only attained in a region
of few G even near the broadest resonances with $\Delta\simeq$50 G.
Outside this region, at least a two-channel model based on
the parameters reported in Tab.~\ref{table1} is needed \cite{goral}.

\section{Outlook} As shown in Tab. \ref{table2},
at least one broad resonance ($\Delta \sim$50 G) is available for
$^{39}$K atoms prepared in each level of the lowest hyperfine
manifold. By virtue of their large width such resonances can be used
to precisely tune the interactions in un ultracold sample. In fact
we have recently demonstrated how to exploit the broad resonance at
402 G in the absolute ground state in order to produce a stable
$^{39}$K Bose-Einstein condensate with widely tunable properties
\cite{nostro}. This system might allow one to study a broad range of
phenomena ranging from atom interferometry with weakly interacting
condensates and strongly-correlated systems in optical lattices to
molecular quantum gases and Efimov physics.

\begin{figure}
\includegraphics[width=\columnwidth,clip]{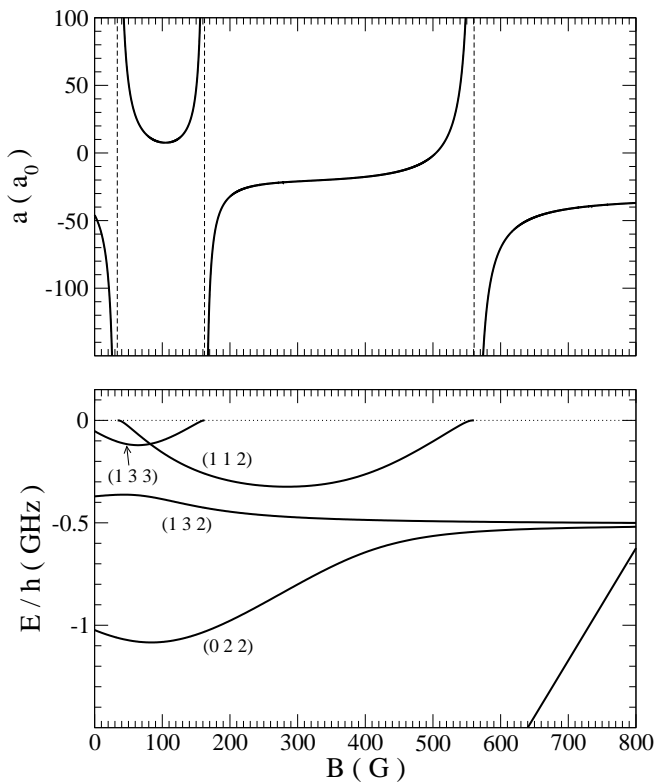}
\caption{Same
as Fig.~\ref{fig3} but for $|1,0\rangle $ atoms and $m_f=0$.
 }
\label{fig4}
\end{figure}

\begin{figure}
\includegraphics[width=\columnwidth,clip]{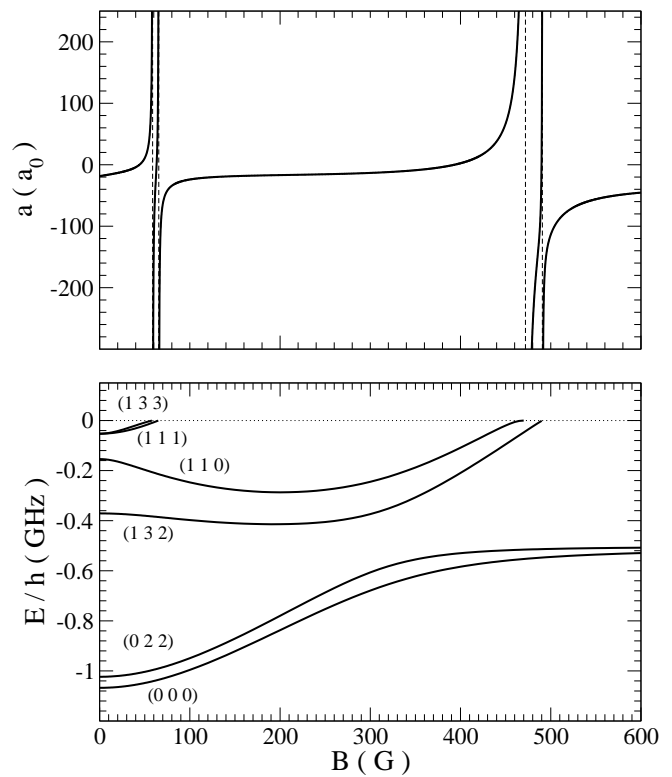}
\caption{Same as Fig.~\ref{fig3} but for $|1,-1\rangle $ atoms and $m_f=-2$.}
\label{fig5}
\end{figure}

The small background scattering length makes this system
particularly appropriate for the exploration of regimes of weak
interactions. At the zero-crossings associated to broad resonances
one can indeed achieve a precise control of $a$ around zero in a
Bose-Einstein condensate. For example, at the zero-crossing location $(350.4 \pm 0.4)$
for $|1,1\rangle + |1,1\rangle$ collisions, the model predicts
a small magnetic-field sensitivity $da$/$dB \simeq$ 0.55 $a_0$/G. This
would imply a control of $a$ to zero within 0.05 $a_0$ for a field
stability of 0.1 G. The other broad resonances for collisions of $|1,0\rangle$ and
of $|1,-1\rangle$ atoms can likewise be used to produce and
manipulate a $^{39}$K condensate. For example, we have already
experimentally verified that a condensate can be produced by
evaporative cooling exploiting either low-field resonance in
$|1,-1\rangle$. Comparing Fig. \ref{fig0} and Fig. \ref{fig4} one
should note how the magnetic-field region around 80 G in which the
scattering length of this state is small and positive ($a\simeq$11
$a_0$) coincides with the maximum in energy of the state. Here an
ultracold $^{39}$K sample would therefore present at the same time
relatively weak interatomic interactions and nearly vanishing
magnetic moment. This peculiar combination is clearly interesting
for interferometric applications and is worth further investigation.

Molecule formation in $^{39}$K condensates can also be studied,
provided a three-dimensional optical lattice is employed to prevent
collapse of the condensate on the atomic side of the resonances and
to shield inelastic decay of molecules \cite{thalhammer}. Also
Feshbach resonances due to molecular states with $\ell\neq$0 are in
principle present in this system, and will be the subject of future
investigation.

Our accurate analysis on $^{39}$K can also be used to calculate the
magnetic-field dependent scattering length of the other bosonic
isotope, $^{41}$K. Bose-Einstein condensation of this species can be
achieved without the need of Feshbach resonances, because of the
naturally positive scattering length \cite{modugno}. Our analysis
shows that a few resonances exist for magnetic fields in the range
0-1000 G, although much narrower than in $^{39}$K. This makes
$^{41}$K less interesting for applications where a precise tuning of
the interactions is needed. Positions and widths of
Feshbach resonances calculated for our best-fit parameters
in the $F$=1 manifold are
reported in Tab. \ref{table3}.
\begin{table}[t]
\begin{center}
\caption{Theoretically calculated
positions $B_{\rm th}$ and widths $\Delta$
of $\ell=0$ Feshbach resonances for $^{41}$K.
Only resonances with $\Delta>10^{-3}$~G are reported.
}
\label{table3}
\vskip 12pt
\begin{tabular}{l | c c }
\hline \hline
 $m_{F},m_{F}$& $B_{\rm th}$(G)&$\Delta_{\rm th}$ (G) \\
 \hline \hline
$1,1$                   & 408.7   & 0.03   \\
                         & 660.1  & 0.2     \\
                         & 856.8  & 0.002   \\
 \hline
$0,0$                   & 451.5  & 0.01      \\
                        & 702.7  &  0.3      \\
                        & 900.1  &  0.002    \\
\hline
$-1,-1$                 & 51.4  & -0.3   \\
                        & 499.9  &  0.004   \\
                        & 747.0  &   0.2  \\
                        & 945.6  &  0.002  \\
 \hline \hline
\end{tabular}
\end{center}
\end{table}

In conclusion, we have presented a detailed analysis of Feshbach
resonances in ultracold $^{39}$K atoms. The full characterization of
collisional parameters, Feshbach resonances and molecular levels we
provide here should be important for planning and interpreting
future experiments with ultracold bosonic potassium. The broad
Feshbach resonances available in $^{39}$K atoms are interesting for
precise control of the interaction in a Bose-Einstein condensate
over a broad range, and for experiments on ultracold molecules.

We thank all the LENS Quantum Gases group for useful discussions and
E. Tiesinga for providing the initial potassium potential
subroutine. M.F. acknowledges support by Centro Enrico Fermi, Roma.
This work was supported by MIUR, by EU under contracts
HPRICT1999-00111 and MEIF-CT-2004-009939, by Ente CRF, Firenze, and
by CNISM, Progetti di Innesco 2005.


\begin{thebibliography}{99}
\bibitem{note heating} The long hold time used for the narrow
resonance explains the absence of heating: after a collisional event
and the subsequent heating, the system has time to evaporate and
rethermalize to a lower temperature.
\end{thebibliography}
\end{document}